# Vapor-deposited non-crystalline phase vs ordinary glasses and supercooled liquids: evidence for significant thermodynamic and kinetic differences.

*Deepanjan Bhattacharya and Vlad Sadtchenko**

The George Washington University

Chemistry Department

Washington, DC

*corresponding author
vlad@gwu.edu

ABSTRACT. Vapor deposition of molecules on a substrate often results in glassy materials of high kinetic stability and low enthalpy[1-6]. The extraordinary properties of such glasses are attributed to high rates of surface diffusion during sample deposition, which makes it possible for constituents to find a configuration of much lower energy on a typical laboratory time scale[1,2,7]. The exact structure of the resulting phase is often assumed to be identical to that obtained by aging of ordinary glass over exceedingly long times. Using Fast Scanning Calorimetry technique, we show that out-of-equilibrium relaxation kinetics and possibly the enthalpy of vapor-deposited films of toluene, an archetypical fragile glass former, are distinct from those of ordinary supercooled phase even when the deposition takes place at temperatures above the glass softening. These observations provide support to the conjecture that the vapor-deposition may result in formation of non-crystalline phase of unique structural, thermodynamic, and kinetic properties.

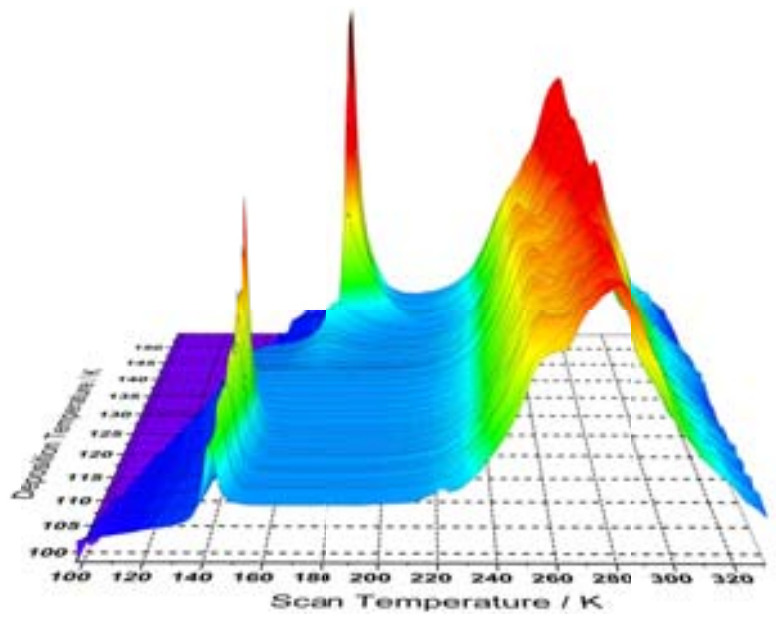

The discovery of highly stable glasses prepared by physical vapor deposition stimulated intense experimental and theoretical efforts to elucidate structure and properties of these remarkable materials[1-21]. In early experiments, the successful preparation of the high stability phase was typically observed when the deposition rates were low and the deposition temperatures, $T_d$, were approximately 0.86 of the $T_g$ [1,3]. It was assumed that, for a particular temperature, the deposition time scale, $t_d$, had to be sufficiently long so that constituents would have sufficient time to diffuse along the sample surface and find a minimum energy configuration before being trapped under subsequent molecular layers[1-3,23].

The empirical scenario of stable phase formation which postulates that vapor deposition simply allows much faster relaxation of the sample to a state, which can be achieved by aging ordinary glasses over extraordinary long times is supported by a number of experimental observations. For example, adiabatic calorimetry studies by Ramos *et. al.* imply that the enthalpy of macroscopic samples of vapor-deposited ethylbenzene is equal to that of the equilibrium liquid at temperatures down to 105 K, *i.e.*, at 0.91 of $T_g$. Dalay *et. al.* showed that the densities of indomethacin samples vapor-deposited at temperatures from 0.9 to 1.03 $T_g$ are consistent with the extrapolated density of ordinary supercooled liquid in the same temperature range. Furthermore, Fakhraai and colleagues demonstrated that the speed of sound in vapor-deposited indomethacin glass is consistent with the values predicted for ordinary supercooled liquid in the similar temperature range.

Despite available experimental evidence in support of structural and thermodynamic similarities between vapor-deposited phases and corresponding aged glasses, there are also several experimental results, which seem to contradict the assumption that the vapor-deposition results in a phase which is simply an ordinary glass characterized by an extraordinary low fictive temperature. Of particular interest are possible experimental observations and computational results which indicate that the vapor-deposition may result in formation of an anisotropic phase. For example, past wide angle X-ray scattering (WAXS) studies revealed structural anisotropy in stable VD tris-naphthylbenzene films[15]. More recent WAXS and computational investigations of stable toluene samples by Ishii and Nakayama also imply that the structure of vapor-deposited phase of toluene may be distinct from the ordinary glass, and may be characterized by presence of locally stable aggregates[6]. Finally, the formation of anisotropic (layered) stable phase was also observed in computational studies of the structure of a VD phase of trehalose[11]. Although relevance of anisotropy of vapor-deposited phase to its stability was questioned[15], the latest low-temperature heat capacity measurements of indomethacin brings the possible difference in structure and physical properties of vapor-deposited and ordinary cooled glasses back into focus[24]. In short, observation of structural anisotropy of vapor-deposited stable phases may imply that, at least in some cases, vapor-deposition may result in non-crystalline phase which structure may be drastically different than those of the corresponding low-temperature equilibrium liquid.

With the objective of clarifying the formation mechanism, molecular structure, and thermodynamic and kinetic properties of VD and OS non-crystalline phases, we initiated a FSC study of selected low temperature organic glasses. The central idea of the FSC technique consists in measuring heat capacity of VD or OS sample during heating from an initial state with rates ranging up to $10^8$ times higher than those employed in traditional Differential Scanning Calorimetry (DSC). The primary advantage of the high heating rate is the high temperature of relaxation during the scan. The return to equilibrium during a rapid scan is manifested in the FSC thermograms by large and sharp fast relaxation endotherms (FREs), which onset temperature, $T_{FRE}$, and the magnitude are highly sensitive to sample preparative conditions[25]. Because the kinetics of high temperature relaxation during fast scan must depend on the initial structural and thermodynamic state of the sample, the FREs can be used to probe the sample properties in a wide range of temperatures from those below and *significantly above* the standard glass transition, $T_g$.

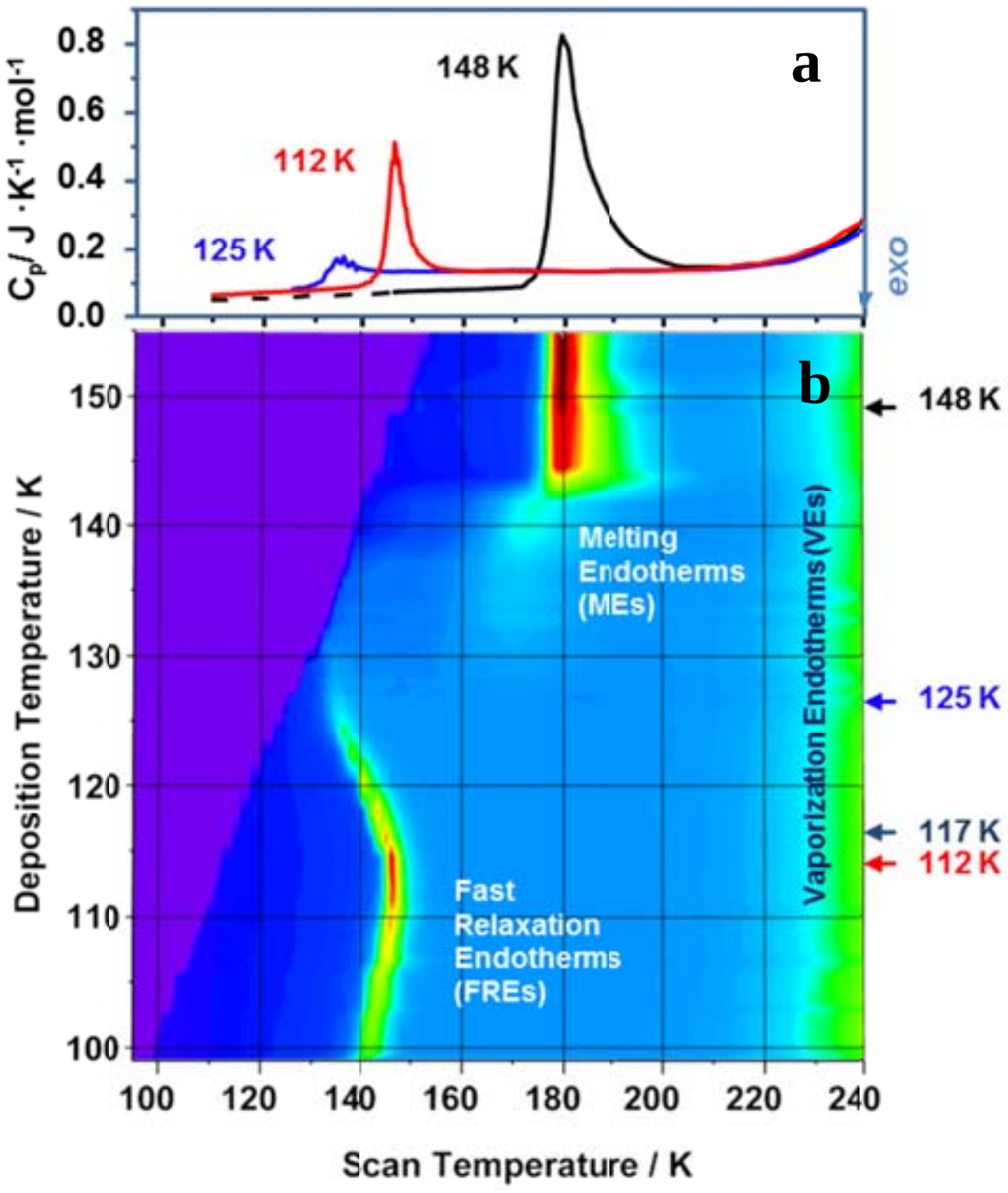

**Figure 1 a,** Representative FSC thermograms of toluene films vapor-deposited at 112, 125, and 148 K. **b,** Toluene "*thermoscape*", i.e., 50 individual thermograms similar to those shown in Fig. 2a., presented in the form of a surface plot of heat capacity as function of deposition and scan temperatures.

Figure 1 summarizes the results of the FSC studies of toluene films vapor-deposited at temperatures between 95 and 155 K. The representative FSC thermograms are shown in Fig. 1a for samples deposited at 112 K, 125 K, and 148 K. 50 thermograms of toluene films deposited at various temperatures were combined into a surface plot termed "thermoscape", which presents the heat capacity of a sample as a function of temperature during scan, and the $T_d$. The results are shown in Fig. 1b. The endotherms are marked by shades of green, yellow and red. The deposition rate in all experiment was approximately 15 nm·s$^{-1}$. Each film was deposited for 120 s, which resulted in film thickness of 1.8 μm. The heating rate in all experiments was approximately $10^5$ K·s$^{-1}$.

As shown in the Fig. 1 b, deposition at 97 K results in FSC thermograms characterized by a FRE near 136 K, and a broad endotherm near 230 K. The endotherms near 230 K are due to the onset of film vaporization as verified by mass-spectrometric measurements. As $T_d$ increases, the FREs gradually shift toward higher temperatures. The increase in the $T_{FRE}$ is accompanied by an increase in the FREs magnitudes. The increase in the $T_{FRE}$s and the endotherm magnitudes is consistent with higher kinetic stability, *i.e.*, with lower transformation rates of VD samples into OS liquid during rapid heating[25].

At $T_d$ near 112 K, the VD films reach the maximum of kinetic stability. Further increase in $T_d$ leads to a rapid decrease in $T_{FRE}$, and a rapid decrease in the FRE's magnitude. When $T_d$ approaches 130 K, the FREs "disappear" from the thermograms. The lack of observable endotherms in the thermograms of toluene deposited at temperatures above 130 K is consistent with its near-equilibrium relaxation times[26,27], $\tau_{rel}$, approaching the characteristic observation time, $t_{FSC}$, of the FSC experiments (approximately $10^{-5}$ s). In other words, heating rates in excess of those used in our experiments ($10^5$ K·s$^{-1}$) are necessary to shift OS liquid sample out of equilibrium at this temperature.

As $T_d$ approaches 148 K, the deposition results in crystallization of toluene samples, which is manifested as large melting endotherms (MEs) near 178 K[22]. Note that unlike the FREs, the onset temperatures and magnitudes of MEs are independent of the deposition conditions. The strong dependence of FREs on the deposition condition, and the lack of such in the case of MEs is a textbook example of fundamental distinctions between glass softening (or liquid relaxation) from typical first order phase transitions[28].

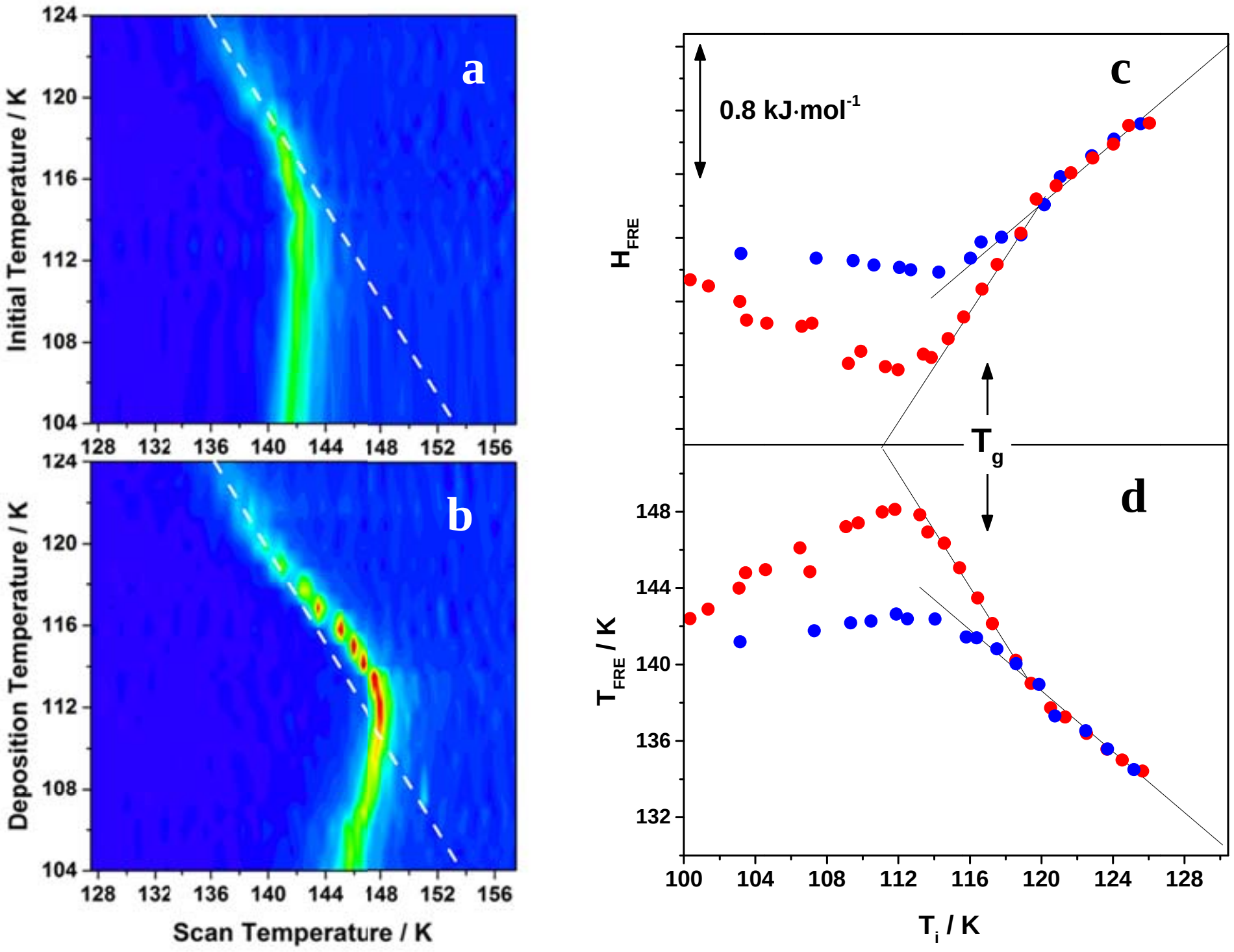

Figure 2 a, FREs of “ordinary” supercooled and glassy toluene films. The dotted line is an eye guide. b, FREs of vapor-deposited toluene films. The dotted lines are eye guides. c, Enthalpy ($H_{FRE}$,) of vapor-deposited and ordinary toluene samples (red and blue dots respectively) just prior to relaxation ,i.e., at $T_{FRE}$ . The enthalpy values are plotted as function of initial (or deposition) temperature of the sample. d, The onset temperatures of relaxation during rapid scans, $T_{FRE}$, plotted as function of the initial temperature for the vapor-deposited and ordinary samples (red and blue dots respectively).

Figures 2a and b compare the thermoscape of VD films with that of OS samples. The OS films were prepared by deposition of toluene at temperatures near 127 K and then cooled to a lower temperature (initial temperature of the FSC scan or $T_i$) with the rate of 3.6 $K \cdot s^{-1}$. As shown in Fig 2, the FREs of VD and OS films are strikingly different. FREs of the OS sample shifts to higher temperatures as $T_i$ is lowered until $T_i$ is near 115 K. Further decrease in $T_i$ does not lead to significant variations in the FRE position on the thermoscape. Such behavior is consistent with kinetic slowdown in ordinary melts upon glass hardening transition[28] in toluene below 117 K. However, in the case of VD samples, the shift of FREs to lower temperatures continues until $T_d$ (in this case, same as $T_i$) is near 112 K. Further decrease in $T_d$ results in a broad maximum in $T_{FRE}$, which is followed by a decrease as $T_d$ is lowered from 110 to 96 K.

Figures 2c and d summarize results of analysis of the thermograms in Fig. 2a and b. They show $T_{FRE}$ and the enthalpy ($H_{FRE}$) of the rapidly heated sample just prior to relaxation (i.e., at $T_{FRE}$). The symmetry in plots of $H_{FRE}$, and $T_{FRE}$ may be interpreted evidence of a strong correlation between kinetic stabilities of the samples and their initial enthalpies[25]. More important for current discussion, however, are the following two observations. *(i)* As shown in Fig. 2, $T_{FRE}$ and $H_{FRE}$ of VD films are still distinct from those of the OS films, at temperatures as high as 119 K, *i.e.*, 2 K above $T_g$ of toluene. *(ii)* The temperature dependence of $H_{FRE}$ of the VD samples on $T_d$ changes slope at 119 K. However, $H_{FRE}$ of OS phase can be approximated by a linear function of $T_i$ in the range from 125 to 115 K. Based on these observations, we provide an argument that the kinetic, thermodynamic, and structural properties of VD phase are distinct from those of OS liquid or glassy samples immediately below.

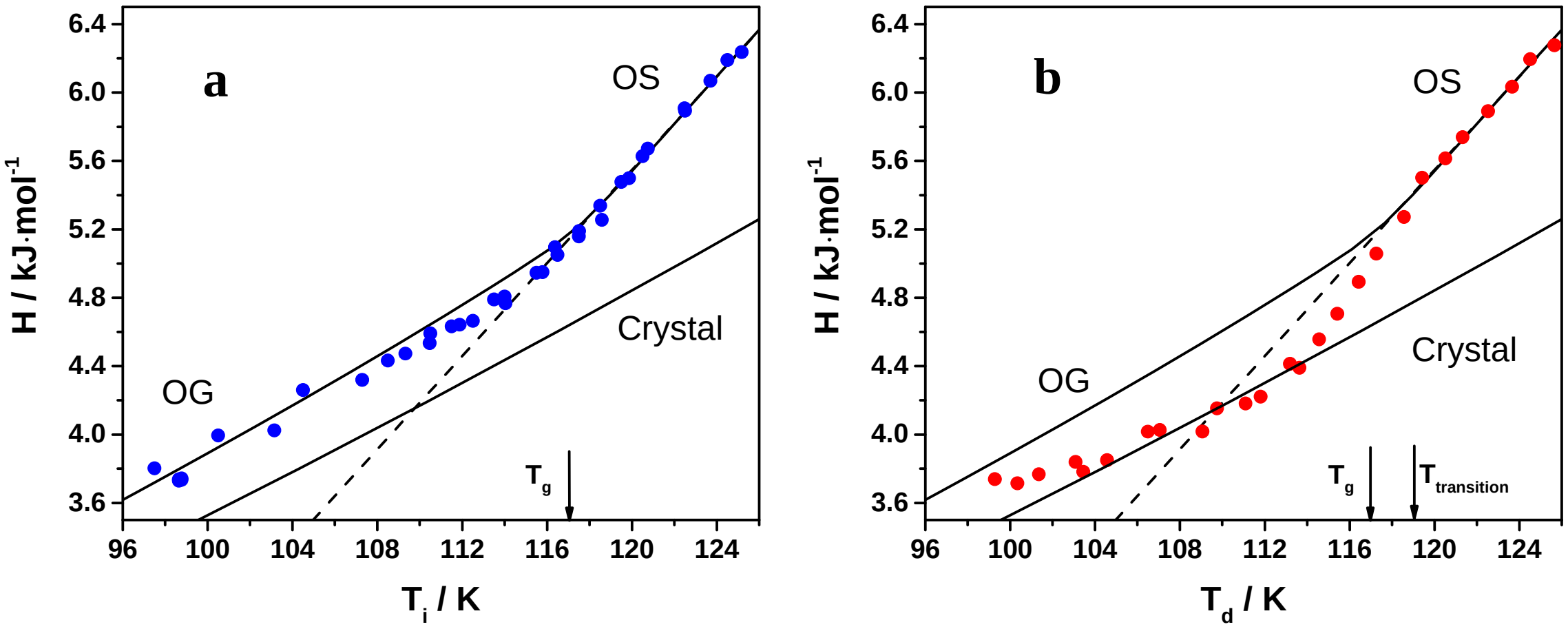

**Figure 3 a,** Estimates of absolute enthalpy, H, of ordinary supercooled liquid and glassy films prior to rapid heating plotted as a function of initial temperature. The solid lines are the enthalpies of ordinary supercooled liquid and glass (marked OS and OG) and the crystalline toluene from DSC studies. T dashed line is the enthalpy of supercooled liquid extrapolated to temperatures below standard $T_g$. **b,** The results of extrapolation of enthalpy values from $H_{FRE}$ at $T_{FRE}$ to lower temperatures using the same approach as that illustrated in Fig. 2a.

Let us assume now that the VD films are structurally, thermodynamically, and kinetically identical to OS liquid or ordinary glass (OG), and that the surface diffusion of the constituents simply makes it possible to accelerate relaxation of VD samples to the equilibrium state. In this case, vapor deposition and traditional cooling of the melt must result in identical phases as long as the characteristic cooling or annealing ($t_a$), and the vapor deposition times ($t_d$) are longer than $\tau_{rel}$ at a particular temperature. $\tau_{rel}$ in supercooled toluene films is less than 5.5 s at 118 K[26,27]. $t_d$ of VD samples was 120 s, and $t_a$ of OS films was 600 s. However, vapor deposition clearly results in formation of films of noticeably higher kinetic stability and possibly lower initial enthalpy at this temperature. Thus, we conclude that vapor-deposition must result in formation of thermodynamically and structurally distinct phase. As we illustrate immediately below, this conclusion is also supported by the observed change in the slope of $H_{FRE}$ dependence on $T_d$ at 119 K.

Fig. 3 shows absolute *initial* (i.e., prior to FSC scan) enthalpies of OS and VD samples inferred from $H_{FRE}$ values and plotted as a function of $T_i$ or $T_d$. Unfortunately, noise in the thermograms made direct determination of absolute enthalpy by integration of the thermograms difficult. Thus, the initial enthalpy values were obtained by extrapolation of $H_{FRE}$ values to the initial temperature of the samples ($T_i$ or $T_d$). The solid lines are the enthalpies of crystalline and ordinary supercooled and glassy toluene inferred from past high accuracy DSC measurements[22,23]. The dashed line is the enthalpy of supercooled liquid extrapolated to temperatures below $T_g$.

As shown in Fig. 3 a, the enthalpy of OS films follows previously established values at temperatures above 115 K. At temperatures below 115 K, the OS toluene gradually falls out of equilibrium, *i.e.*, undergoes the laboratory glass transition. Note that deviation from equilibrium occurs in our experiments at 115 K and not at 117 K as observe in the DSC measurements[22]. This is due to the fact that the OS samples in our experiments *were aged* at 115 K for 600 s prior to the FSC scan to ensure that the enthalpy of OS phase was *not* higher than that of the OS samples used in past DSC investigations. In short, the enthalpy values in Fig. 3a are characteristic of OS phase fully equilibrated at 115 K. Note that, despite aging, the OS phase never crosses the enthalpy line of equilibrium supercooled liquid (dashed line). A drop in enthalpy below equilibrium values of supercooled liquid may signify a phase transition as, for example, in the case of crystallization.

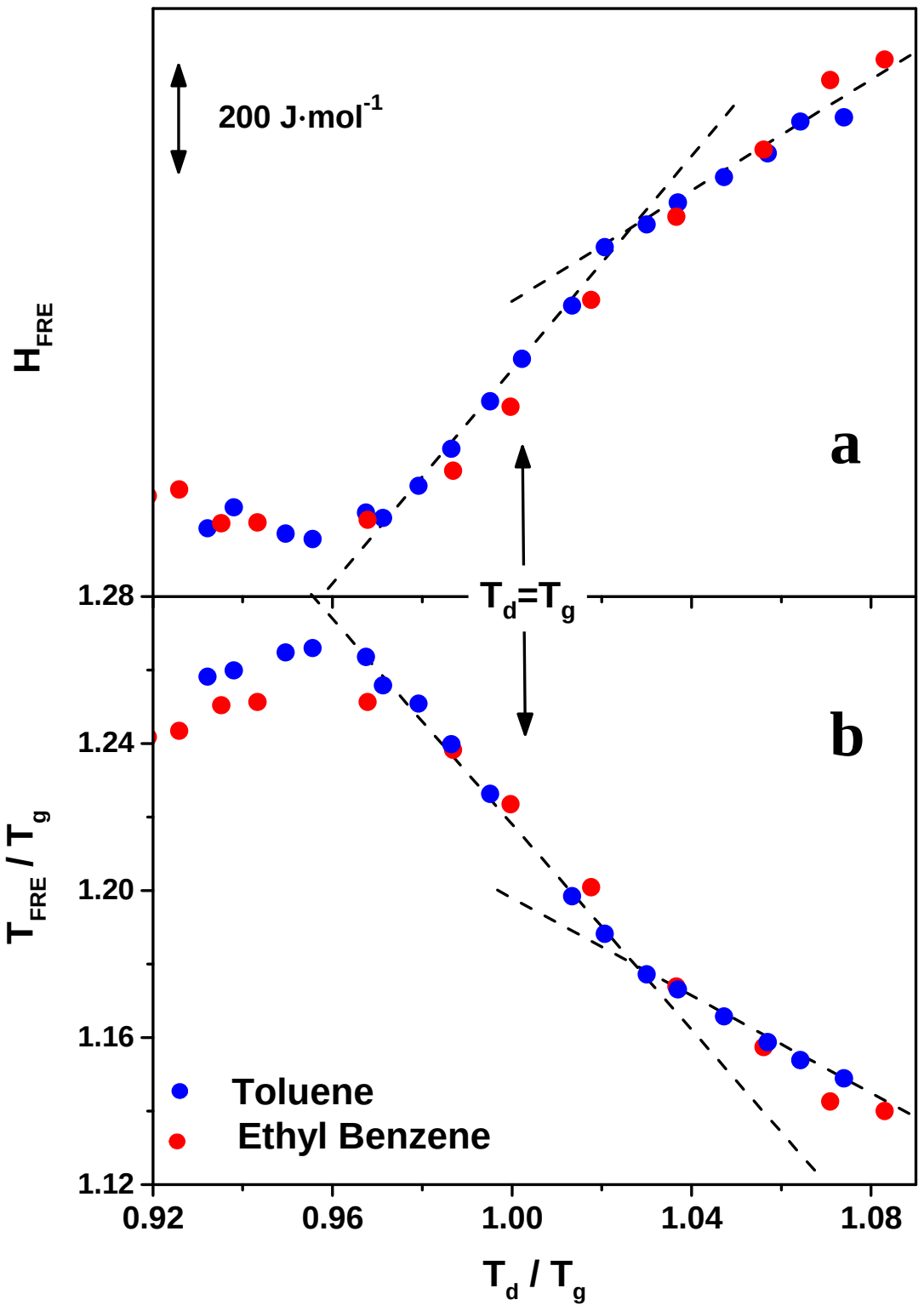

**Figure 4. a,** Enthalpy ($H_{FRE}$,) of vapor-deposited toluene and ethylbenzene samples (blue and red dots respectively) just prior to relaxation ( at $T_{FRE}$). The enthalpy values are plotted as function of deposition temperatures scaled by the standard $T_g$s of toluene and ethylbenzene. **b**, *Scaled* onset temperatures of relaxation during rapid scans ($T_{FRE}$/$T_g$) plotted as functions of *scaled* deposition temperatures ($T_d$/$T_g$) in the case of VD toluene (blue dots) and ethylbenzene films (red dots).

As shown in Fig. 3b, the enthalpy of VD toluene is identical to that of the OS liquid when deposition temperature is above 119 K. At temperatures below 119 K, however, the enthalpy dependence on temperature changes its slope resulting in values progressively lower than those of the OS phase (dashed line). Note that this first change in the slope of the enthalpy dependence on temperature is followed by the second change at 112 K, which coincides with the samples achieving the maximum kinetic stability. *Note that*, at *this point, the enthalpy of VD toluene samples becomes equal or ever slightly lower than the enthalpy of the crystalline phase.* In short, the variations in the enthalpy of VD toluene films with deposition temperature are inconsistent surface diffusion accelerated relaxation of the VD sample into an ordinary low temperature liquid state. In fact the complex enthalpic path and low enthalpy values imply formation of thermodynamically and structurally distinct phase. Before we speculate on the structure and properties of VD phases, we would like to refute one potential concerns with the validity of our conclusions.

It can be argued that the temperature calibration of our apparatus is incorrect, *i.e.*, the actual $T_i$ or $T_d$ values are 2 K lower than those in Fig. 2 and 3, and that $H_{FRE}$ and $T_{FRE}$ of OS and VD samples deviate only when $\tau_{rel}$ is longer than the annealing and deposition times, i.e., at temperatures below $T_g$.  We emphasize, however,  that the distinctions in $T_{FRE}$ and $H_{FRE}$ of OS and VD samples cannot be explained simply by uncertainties in temperature determination, and not just due to our extensive tests and calibration procedures[25].

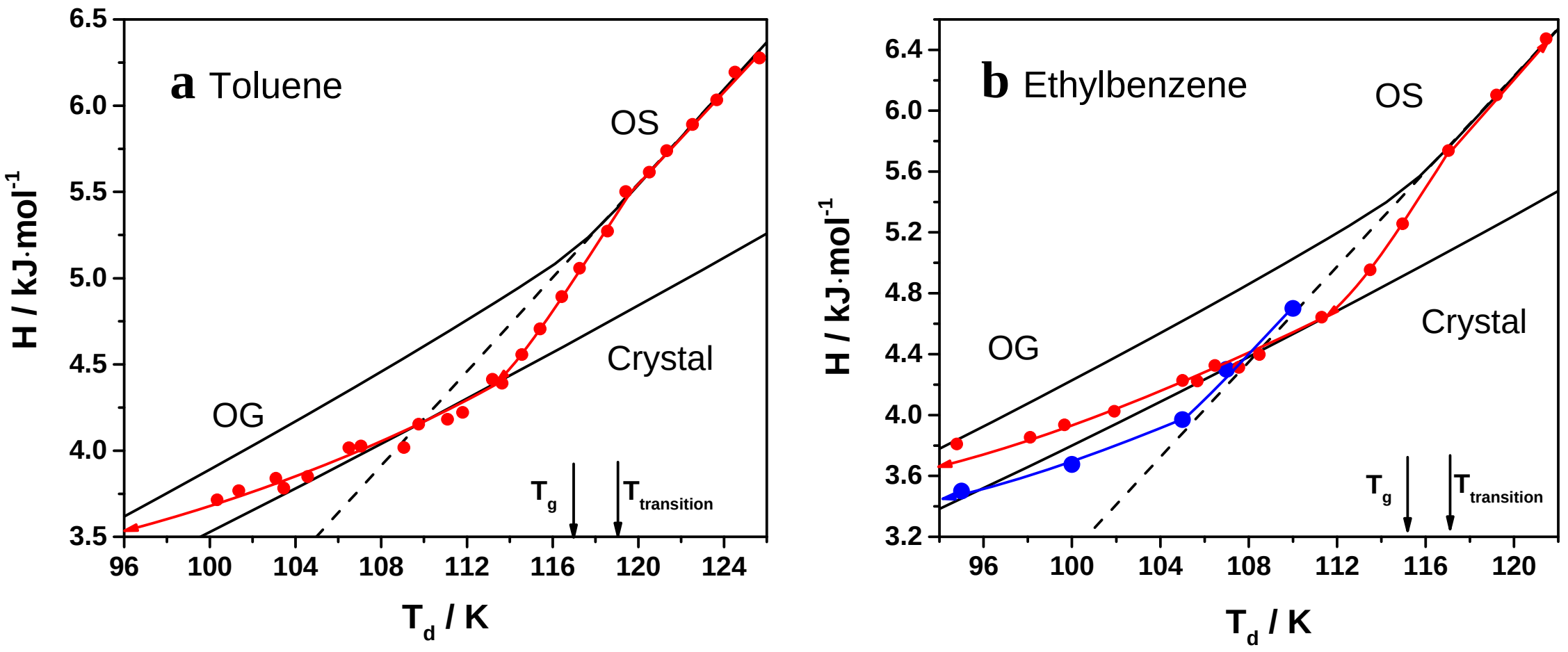

**Figure 5** Estimates of absolute enthalpy, H, of toluene (a) and ethylbenzene (b) vapor-deposited different temperatures. The solid lines are the enthalpies of ordinary supercooled liquid and glass (marked OS and OG) and the crystalline phases from DSC studies[22]. The dashed lines are the enthalpies of supercooled liquids extrapolated to temperatures below $T_g$s. Blue and red curves connecting the data points are the eye guides. The large blue dots and the corresponding curve represent the results of adiabatic calorimetry stud by Ramos *et. al*

As shown in Fig. 3a, the enthalpies of OS samples prepared by cooling of the melt are nearly equilibrated at temperatures as low as 115 K. The equilibration of the OS samples at temperature below $T_g$ was possible due to 10 min. aging at 115 K. Note, however, that such a relatively short equilibration time is *impossible* at temperature below 115 K. Indeed $\tau_{rel}$ at 115 K is already near 900 s and is expected to decrease by an order of magnitude at 114 K[26,27].  In short, the deposition and starting temperatures reported in in this article represent *the lowest possible estimates* of the actual values.

In order to demonstrate that "anomalous" dependence of kinetic and thermodynamic variables of VD non-crystalline phase is not limited to the case of toluene, we collected thermograms of ethylbenzene (EB) vapor-deposited at distinct temperatures. Fig. 4a and b compare the dependences of $T_{FRE}$ and $H_{FRE}$ on deposition temperature in the case of toluene and EB samples prepared under identical experimental conditions. Note that the standard glass transition temperature of EB is 2.5 K lower than that of toluene, which resulted in a dependence of $T_{FRE}$ on $T_d$ distinct from that in the case of toluene. Intriguingly, as shown in Fig. 4b, when the $T_d$ and $T_{FRE}$ from both sets of experiments were normalized (divided) by the standard $T_g$s of two compounds, the $T_{FRE}$ dependences collapsed into nearly identical plots. Although the meaning and significance of this observation are not understood at this time, Fig. 4b clearly shows that dependence of $T_{FRE}$ on $T_d$ in the case of EB also changes slope a few degrees above the standard $T_g$. As shown in Fig. 4a, $H_{FRE}$ dependence on $T_d/T_g$ parameter in the case of vapor-deposited EB samples is also very similar to that in the case of vapor deposited toluene films (the $H_{FRE}$ values were offset to facilitate comparison).

Figure 5 compares the absolute initial enthalpies of VD toluene and ethylbenzene. As shown in the figure in both cases, the enthalpy of VD phase begins to decrease below OS liquid enthalpy at temperatures a few degrees above $T_g$. In both cases, the maximum kinetic stability is achieved when the enthalpy of the VD phase reaches values near or slightly lower than those of crystalline phase. In short, two glass formers of distinct molecular structure investigated in our FSC experiments demonstrate essentially the same "anomalous" dependence of kinetic and thermodynamic parameters, which indicates that such an anomalous behavior may be a general a general phenomenon.

Finally, we compare results of our FSC studies of vapor-deposited EB samples with those from adiabatic calorimetry (ADC) experiments[13]. The ADC data are shown in Fig. 5b as large blue dots. Taking into account dramatically differences in VD sample preparation conditions (e.g., 5 times lower deposition rate and 500 times greater film thickness in the ADC experiments), and the differences in the sample interrogation methods ($10^9$ times higher heating rates in the FSC experiments), the similarities in the temperature dependences of enthalpy dependence on temperatures are striking. Indeed, in both cases, a decrease in enthalpy occurs at some temperature over a range of 4 to 6 K as $T_d$ is lowered. In both cases, the initial drop in the enthalpy is followed by a gradual decrease at lower temperatures along a path which is unlike that of ordinary glass or crystal. In both cases, the highest kinetic stability is achieved when the enthalpy of the VD sample is equal or *even significantly lower* than the enthalpy of the crystalline phase. In short, in both cases, the dependence of enthalpy of VD phase on temperature is inconsistent with transition of supercooled liquid into an ordinary glass.

In summary, the results of the FSC experiments presented in this letter provide strong support to the conjecture that the vapor-deposition may result in formation of unique non-crystalline phase. What is likely structure of such a phase or phases? At this point we can only speculate. Based on the results of obtained in our recent FSC experiments [25], we suggested that VS samples represent a stacked phase consisting of 2D islands of lower enthalpy and higher kinetic stability than those of ordinary glass. Future experiments are required, of cause, to elucidate the structure of highly stable non-crystalline material prepared by vapor deposition.

**Materials and methods**

The experiments were conducted using custom-built fast scanning calorimeter[25]. The "sample holder" is essentially a 10 μm in diameter, 1.5 cm long tungsten filament attached to supports inside a vacuum chamber maintained at a pressure of approximately $5 \cdot 10^{-7}$ Torr. The temperature of the filament during film's preparation is controlled by adjusting the temperatures of the supports. Films are deposited on the surface of the filament via 12 effusive dosers. After film preparation, heating of the filament is initiated by applying a potential difference across the filament. The data acquisition system simultaneously measures the voltages drop across the filament and the current through the filament. These data are used later to calculate the filament resistance and the dissipated power. The temperature of the filament is inferred from its resistance using results of a calibration procedure[25]. Heat capacity of the filament is calculated as the ratio of the power dissipated by the filament to the first time derivative of its temperature.

**Acknowledgments:** This work was supported by the US National Science Award CHE-1012692. We are grateful to Liam O'Reilly for assistance with some measurements. We are also grateful to Ulyana S. Cubeta for her tests and analysis of the temperature calibration procedure used in the FSC experiments.

## References

(1) Swallen, S. F.; Kearns, K. L.; Mapes, M. K.; Kim, Y. S.; McMahon, R. J.; Ediger, M. D.; Wu, T.; Yu, L.; Satija, S. *Science* **2007**, *315*, 353.

(2) Kearns, K. L.; Swallen, S. F.; Ediger, M. D.; Wu, T.; Sun, Y.; Yu, L. *Journal of Physical Chemistry B* **2008**, *112*, 4934.

(3) Dawson, K. J.; Kearns, K. L.; Yu, L.; Steffen, W.; Ediger, M. D. *Proceedings of the National Academy of Sciences* **2009**, *106*, 15165.

(4) Ediger, M. D.; Harrowell, P. *Journal of Chemical Physics* **2012**, *137*, 080901.

(5) Singh, S.; Ediger, M. D.; de Pablo, J. J. *Nature Materials* **2013**, *12*, 139.

(6) Ishii, K.; Nakayama, H. *Physical Chemistry Chemical Physics* **2014**, *16*, 12073.

(7) Daley, C. R.; Fakhraai, Z.; Ediger, M. D.; Forrest, J. A. *Soft Matter* **2012**, *8*, 2206.

(8) Kearns, K. L.; Ediger, M. D.; Huth, H.; Schick, C. *Journal of Physical Chemistry Letters* **2010**, *1*, 388.

(9) Leon-Gutierrez, E.; Sepulveda, A.; Garcia, G.; Clavaguera-Mora, M. T.; Rodriguez-Viejo, J. *Physical Chemistry Chemical Physics* **2010**, *12*, 14693.

(10) Chen, Z.; Richert, R. *Journal of Chemical Physics* **2011**, *135*.

(11) Singh, S.; de Pablo, J. J. *The Journal of Chemical Physics* **2011**, *134*.

(12) Sun, Y.; Zhu, L.; Kearns, K. L.; Ediger, M. D.; Yu, L. A. *Proceedings of the National Academy of Sciences of the United States of America* **2011**, *108*, 5990.

(13) Ramos, S.; Oguni, M.; Ishii, K.; Nakayama, H. *Journal of Physical Chemistry B* **2011**, *115*, 14327.

(14) Zhu, L.; Brian, C. W.; Swallen, S. F.; Straus, P. T.; Ediger, M. D.; Yu, L. *Physical Review Letters* **2011**, *106*.

(15) Dawson, K.; Kopff, L. A.; Zhu, L.; McMahon, R. J.; Yu, L.; Richert, R.; Ediger, M. D. *Journal of Chemical Physics* **2012**, *136*.

(16) Guo, Y.; Morozov, A.; Schneider, D.; Chung, J. W.; Zhang, C.; Waldmann, M.; Yao, N.; Fytas, G.; Arnold, C. B.; Priestley, R. D. *Nat Mater* **2012**, *11*, 337.

(17) Ahrenberg, M.; Chua, Y. Z.; Whitaker, K. R.; Huth, H.; Ediger, M. D.; Schick, C. *Journal of Chemical Physics* **2013**, *138*.

(18) Chen, Z.; Sepulveda, A.; Ediger, M. D.; Richert, R. *Journal of Chemical Physics* **2013**, *138*.

(19) Douglass, I.; Harrowell, P. *Journal of Chemical Physics* **2013**, *138*.

(20) Lyubimov, I.; Ediger, M. D.; de Pablo, J. J. *Journal of Chemical Physics* **2013**, *139*.

(21) Sepulveda, A.; Swallen, S. F.; Ediger, M. D. *Journal of Chemical Physics* **2013**, *138*.

(22) Yamamuro, O.; Tsukushi, I.; Lindqvist, A.; Takahara, S.; Ishikawa, M.; Matsuo, T. *Journal of Physical Chemistry B* **1998**, *102*, 1605.

(23) Scott, D. W.; Guthrie, G. B.; Messerly, J. F.; Todd, S. S.; Berg, W. T.; Hossenlopp, I. A.; McCullough, J. P. *The Journal of Physical Chemistry* **1962**, *66*, 911.

(24) Pérez-Castañeda, T.; Rodríguez-Tinoco, C.; Rodríguez-Viejo, J.; Ramos, M. A. *Proceedings of the National Academy of Sciences* **2014**, *111*, 11275.
(25) Bhattacharya, D.; Sadtchenko, V. *Journal of Chemical Physics* **2014**, 008433.
(26) Hinze, G.; Sillescu, H. *Journal of Chemical Physics* **1996**, *104*, 314.
(27) Hinze, G.; Diezemann, G.; Sillescu, H. *Journal of Chemical Physics* **1996**, *104*, 430.
(28) Hodge, I. M. *Journal of Non-Crystalline Solids* **1994**, *169*, 211.